\documentclass[sigconf]{acmart}
\usepackage{xcolor, physics, soul}
\usepackage{amsmath,amsfonts,mathtools}
\usepackage{adjustbox}
\AtBeginDocument{%
  \providecommand\BibTeX{{%
    \normalfont B\kern-0.5em{\scshape i\kern-0.25em b}\kern-0.8em\TeX}}}

\setcopyright{acmlicensed}
\copyrightyear{2018}
\acmYear{2018}
\acmDOI{XXXXXXX.XXXXXXX}

\acmConference[DAC '24]{Make sure to enter the correct
  conference title from your rights confirmation emai}{June,
  2024}{San Francisco, CA}
\acmISBN{978-1-4503-XXXX-X/18/06}

\begin{document}

\title{AltGraph: Redesigning Quantum Circuits Using Generative Graph Models for Efficient Optimization}


% \author{\IEEEauthorblockN{Collin Beaudoin}
% \IEEEauthorblockA{\textit{Penn State University} \\
% % \textit{Department of Computer Science \& Engineering}\\
%  University Park, Pennsylvania, USA \\
%  cpb5867@psu.edu}
%  \and
%  \IEEEauthorblockN{Koustubh Phalak}
%  \IEEEauthorblockA{\textit{Penn State University} \\
%  %\textit{Department of Computer Science \& Engineering}\\
%  University Park, Pennsylvania, USA \\
%  krp5448@psu.edu}
%  \and
%  \IEEEauthorblockN{Swaroop Ghosh}
%  \IEEEauthorblockA{\textit{Penn State University} \\
% % \textit{Department of Computer Science \& Engineering}\\
% University Park, Pennsylvania, USA \\
% szg212@psu.edu}}

\author{Collin Beaudoin}
\email{cpb5867@psu.edu}
\affiliation{%
  \institution{Penn State University}
  % \streetaddress{P.O. Box 1212}
  \city{University Park}
  \state{Pennsylvania}
  \country{USA}
  \postcode{16801}
}

\author{Koustubh Phalak}
\email{krp5448@psu.edu}
\affiliation{%
  \institution{Penn State University}
  % \streetaddress{P.O. Box 1212}
  \city{University Park}
  \state{Pennsylvania}
  \country{USA}
  \postcode{16801}
}

\author{Swaroop Ghosh}
\email{szg212@psu.edu}
\affiliation{%
  \institution{Penn State University}
  \city{University Park}
  \state{Pennsylvania}
  \country{USA}
  \postcode{16801}
}

% \affiliation{%

% \affiliation{%

% \renewcommand{\shortauthors}{Beaudoin and Phalak, et al.}

\begin{abstract}
Quantum circuit transformation aims to produce equivalent circuits while optimizing for various aspects such as circuit depth, gate count, and compatibility with modern Noisy Intermediate Scale Quantum (NISQ) devices. There are two techniques for circuit transformation. The first is a rule-based approach that greedily cancels out pairs of gates that equate to the identity unitary operation. Rule-based approaches are used in quantum compilers such as Qiskit, t$\ket{ket}$, and Quilc. The second is a search-based approach that tries to find an equivalent quantum circuit by exploring the quantum circuits search space. Search-based approaches typically rely on machine learning techniques such as generative models and Reinforcement Learning (RL). In this work, we propose \emph{AltGraph}, a novel search-based circuit transformation approach that generates equivalent quantum circuits using existing generative graph models. We use three main graph models: DAG Variational Autoencoder (\emph{D-VAE}) with two variants: Gated Recurrent Unit (GRU) and Graph Convolutional Network (GCN), and Deep Generative Model for Graphs (\emph{DeepGMG}) that take a Direct Acyclic Graph (DAG) of the quantum circuit as input and output a new DAG from which we reconstruct the equivalent quantum circuit. Next, we perturb the latent space to generate equivalent quantum circuits some of which may be more compatible with the hardware coupling map and/or enable better optimization leading to reduced gate count and circuit depth. AltGraph achieves on average a 37.55\% reduction in the number of gates and a 37.75\% reduction in the circuit depth post-transpiling compared to the original transpiled circuit with only 0.0074 Mean Squared Error (MSE) in the density matrix.
\end{abstract}

\settopmatter{printacmref=false}
\renewcommand\footnotetextcopyrightpermission[1]{}
\pagestyle{plain}




% \received{20 November 2023}

\maketitle

\section{Introduction}
Quantum circuits are the fundamental component of modern-day quantum computing. They are analogous to classical electronic circuits, where various classical (quantum) gate operations are performed on bits (qubits) to obtain the desired output. Typically, each quantum circuit has a set of quantum gate operations followed by a measurement operation on desired qubits at the end. Special quantum circuits, such as Parametric Quantum Circuits (PQC), also have a state preparation component at the start to prepare the initial state of qubits. Each quantum gate is a unitary operation that can be represented as unitary matrix $U$ (for conjugate transpose $U^{*}$, $UU^{*}=UU^{-1}=I$) and works on either a single qubit or multiple qubits. A few well-known single qubit gates are Hadamard (H), Pauli X/Y/Z gates, rotation gates like RX/RY/RZ and U, and multi-qubit gates are CNOT, Toffoli, controlled Pauli and controlled rotation gates.  

Ideal quantum circuits with no noise produce correct output with 100$\%$ fidelity. Unfortunately, NISQ computers have gate errors, readout errors, decoherence errors, crosstalk errors, and other noise sources that degrade the overall fidelity and correctness of computation. In such noisy environments, gate count and circuit depth are salient when crafting a circuit. Typically, circuits with low depths and gate counts are ideal for maximum fidelity. Therefore, generating equivalent quantum circuits to obtain lower-gate functionally equivalent quantum circuits is a fundamental research problem in the NISQ era.

There are two methods of performing \emph{quantum circuit transformation} (referred to as  \emph{circuit transformation} for brevity) \cite{li2023quarl} namely, \emph{rule-based} and \emph{search-based}. The former uses gate-cancellation rules to optimize quantum circuits\cite{anis2021qiskit}, while the latter uses machine learning techniques to find functionally equivalent quantum circuits from the search space of quantum circuits \cite{li2023quarl,schulman2017proximal,fosel2021quantum,xu2022quartz,zhou2020monte}. These techniques primarily use Reinforcement Learning (RL) methods e.g., \cite{moflic2023graph} uses an RL agent to determine the correct set of gate cancellations to reach optimal depth on ICM+H circuits. Since ICM+H circuits contain only CNOT and Hadamard gates which are easy to optimize, this technique may not necessarily work on quantum circuits involving other gates. \cite{fosel2021quantum} overcomes this issue by using parametric quantum circuits but require circuits to be hardware compliant prior to performing optimization. \cite{xu2022quartz} uses generative methods by generating circuits as Quantum Assembly Language (QASM) instructions, however the search space of the generation algorithm is limited and also does not sample new circuit representations that may fundamentally address the issues like hardware constraint. \color{black} We address these shortcomings by proposing \emph{AltGraph}: usage of generative graph models that produce an equivalent Directed Acyclic Graph (DAG) of the original quantum circuit DAG from which we reconstruct the corresponding quantum circuit. We perturb the latent space of the models to generate equivalent quantum circuits some of which may be more compatible with the hardware coupling map and/or enable better optimization leading to compact circuits. We use multiple graph generation models, specifically DAG Variational Autoencoder (D-VAE) Gated
Recurrent Unit (GRU), D-VAE Graph Convolutional Network (GCN)
and Deep Generative Model for Graphs (DeepGMG) %\emph{D-VAE GRU}, \emph{D-VAE GCN}, and \emph{DeepGMG} 
and perform comparative analysis.

\textbf{Novelty}: The D-VAE and DeepGMG models were able to process DAGs and generate stylistically similar graphs as the original DAGs. However, there was no way to guarantee a reconstructed DAG could be interpreted as a valid quantum circuit. We extend D-VAE and DeepGMG's graph reconstruction capabilities to generate always valid quantum circuits. Furthermore, we enable the capability to generate novel quantum circuits equivalent to the original quantum circuit by sampling from the latent space of the generative models. Finally, we perform detailed comparative analysis of the effectiveness of the generative models in producing compact representation of the original circuits.
%(a) introduce quantum circuit DAG generation restrictions such as equal indegree and outdegree of quantum gate nodes, matching qubit count in edges and enforcing same number of measurment operations as the original circuit; overall, this leads to forcing all DAG generations to be valid quantum circuit DAGs, (b) remove teacher forcing, requiring the model to learn the entire quantum circuit DAG representation, (c) introduce quantum circuit evaluation based on density matrices.
\emph{To the best of our knowledge, this is the first paper that uses generative graph model-based circuit transformation to produce optimized quantum circuits.} 
% \hl{mention in discussion section-- This work is intended to be a proof-of-concept work that does not rely on quantum hardware compliance, such as coupling map and native gate set, and focuses on non-parametric quantum circuits.}
%We intend to address these factors in a follow-up work where a protoype of this idea will be presented.

The rest of the paper's organization is as follows: Section \ref{bg_rw} illustrates the background information required to understand the paper and related work. Section ~\ref{altgraph} presents the approach to convert the models to valid quantum circuit graph generators. Section ~\ref{results} contains visualizations of each model's abilities to process quantum circuits. Section ~\ref{discussion} analyzes the results, and Section ~\ref{conclusion} summarizes the work.

\section{Background and Related Works} \label{bg_rw}

\subsection{Fundamentals of Quantum Computing}
\subsubsection{Qubits} Qubits are the basic elements of a quantum computer, similar to bits in classical computing. Typically, a qubit is denoted by the quantum state
$\ket{\psi}=$
$\big[\begin{smallmatrix}
\alpha \\
\beta
\end{smallmatrix}\big]$. Here, $\alpha$ and $\beta$ are complex values. The square magnitude of $\alpha$, $|\alpha|^2$, indicates the likelihood of the qubit being observed as a classical 0, while $|\beta|^2$ shows the chance of it being observed as a classical 1. Their combined probabilities always sum up to 1, i.e., $|\alpha|^2+|\beta|^2=1$. Every qubit has two primary states: $\ket{0}$ (with $\alpha=1$ and $\beta=0$) and $\ket{1}$ (with $\alpha=0$ and $\beta=1$). Qubits can be manipulated using unitary gates, and when measured, they settle into either a 0 or 1 state.

\subsubsection{Quantum Gates} Quantum gates, which are unitary matrix functions, modify the states of one or multiple qubits. Dependent on qubit technology, gates are implemented in various ways. For instance, superconducting qubits utilize microwave pulses \cite{krantz2019quantum}, while trapped ion and quantum dots employ laser pulses \cite{benhelm2008towards,nguyen2012optically}. On the other hand, Nuclear Magnetic Resonance (NMR) qubits use radio frequency pulses. The speed at which these gates operate can differ significantly based on the technology, with photonic qubits working in picoseconds and NMR qubits taking up to several seconds \cite{ladd2010quantum}. Gates that affect a single qubit encompass the X (NOT) gate, H (Hadamard) gate, and rotational gates like $R_X$/$R_Y$/$R_Z$, and the $U$ gate. Gates impacting multiple qubits consist of the CNOT (controlled-NOT) gate, Toffoli gate, controlled rotation gates, and the Peres gate \cite{peres1985reversible}.

\subsubsection{Quantum circuit} A quantum circuit is a structured series of gate tasks carried out chronologically. It starts with setting up the initial state of the qubits, known as state initialization. Following this, gate operations within the circuit modify qubits to achieve a specific state. The process concludes with a measurement using a particular gate. These quantum activities occur in a quantum Hilbert space \cite{von2018mathematical}. 

\subsubsection{Transpilation} Before running a quantum circuit, more complex gates, like Tofolli, multi-controlled NOT gate, etc. are broken into sets of gates that the quantum device natively understands. Converting gates to hardware native gates is known as \emph{decomposition}. NISQ devices have restricted connectivity i.e. not every qubit is connected to every other qubit. Physical qubit connections are represented in a graph called a coupling map. Because not all qubits are connected, compilation requires mapping logical qubits to physical qubits, termed as \emph{qubit mapping}. Decomposition combined with qubit mapping is formally known as transpilation. If a gate operation is required between qubits that are not connected, swap operations will be required to bring them to physically connected qubits. Each swap operation introduces 3 CNOT gates which increases the circuit depth and gate count. 
\subsection{Quantum Circuit Transformation}

There are two types of circuit transformation: rule-based approach and search-based approach. A rule-based approach involves utilizing gate-cancellation rules on particular subcircuits to reduce them into smaller amounts of gates with lower depths. For example, two back-to-back Pauli X gates computationally equate to identity operation (I) as 
$\begin{bsmallmatrix}
  0 & 1\\
  1 & 0
\end{bsmallmatrix}
\begin{bsmallmatrix}
    0 & 1\\
    1 & 0
\end{bsmallmatrix}
=
\begin{bsmallmatrix}
    1 & 0\\
    0 & 1
\end{bsmallmatrix}$. Similar behavior exists for other single-qubit gates and multi-qubit gates. This kind of circuit transformation is typically incorporated in quantum compilers like Qiskit \cite{anis2021qiskit}. Search-based approaches offer flexibility in terms of optimization. Rather than relying on gate-cancellation rules, this approach searches for a functionally equivalent circuit. Typically, Machine Learning (ML) techniques such as Reinforcement Learning (RL), generative models, or other methods aid the approach. In RL-based search, the current quantum circuit is the state, and the original version of the circuit is the environment. The circuit transformation performed in each step is the action, and based on some input, the agent performs the most rewarding action. Some RL-based works include \raisebox{0.5pt}{\textcircled{\raisebox{-0.9pt}{1}}} \cite{moflic2023graph}, that uses D-VAE to encode a quantum circuit DAG and send that encoding to an RL agent which then decides what action to perform among the possible set of updates, \raisebox{0.5pt}{\textcircled{\raisebox{-0.9pt}{2}}} \cite{li2023quarl}, that uses Hierarchical Advantage Learning (HAL) combined with Proximal Policy Optimization (PPO) \cite{schulman2017proximal} to perform reinforcement learning, and \raisebox{0.5pt}{\textcircled{\raisebox{-0.9pt}{3}}} \cite{fosel2021quantum} that also uses PPO along with Advantage Actor Critique (AAC) scheme-based agent. In contrast, generative approaches such as \cite{xu2022quartz} generate quantum circuits as a Quantum Assembly Language (QASM) sequence of instructions. Other methods include \raisebox{0.5pt}{\textcircled{\raisebox{-0.9pt}{1}}} \cite{zhou2020monte} uses Monte Carlo tree search framework to generate hardware compliant quantum circuits, and \raisebox{0.5pt}{\textcircled{\raisebox{-0.9pt}{2}}} \cite{duncan2020graph} performs ZX-calculus on graphs to perform circuit transformation.

% \begin{figure}[t]
% \begin{adjustbox}{width=\linewidth}
%   \begin{minipage}{\linewidth}
%      \centering
%      \includegraphics[width=.3\textwidth]{figs/test_graph_id0.pdf}     
%   \end{minipage}\hfill
%   \end{adjustbox}
%   \caption{Sample quantum circuit graph.} \label{fig:sample_graph}
% % \end{center}
% \end{figure}

\begin{figure}[t]
    \vspace{-2mm}
     \centering
         \includegraphics[width=0.75\linewidth]{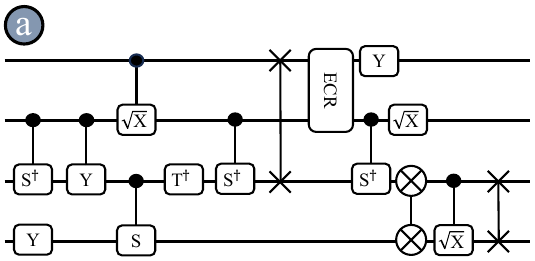}
         \includegraphics[width=0.75\linewidth] {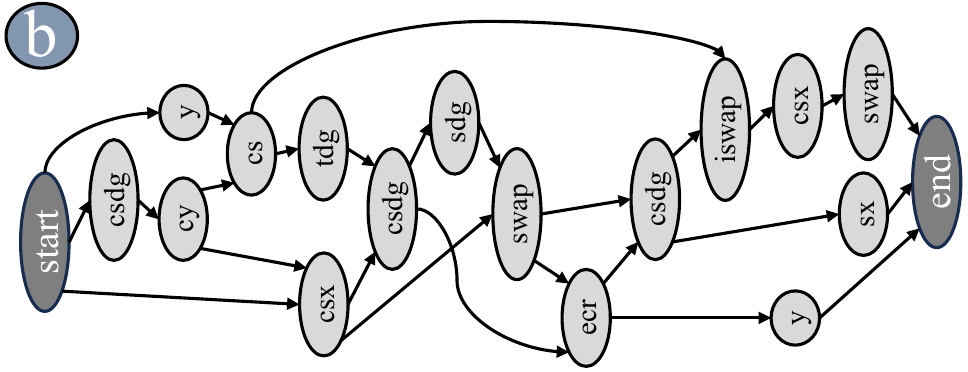} 
     % \vspace{-2mm}
        \caption{(a) Sample 4 qubit 16 gate quantum circuit with (b) the DAG rendering of the circuit.
        }
        \label{fig:sample_circ_graph}
    \vspace{-4mm}
\end{figure}

\subsection{Message Passing}
Generative graph models and graph neural networks, in general, require nodes to share information to build either a global or a local understanding of themselves and the nodes around them in hopes of generating a (semi-)unique representation. While there are multiple techniques, only two approaches exist in D-VAE and DeepGMG: simultaneous message passing and asynchronous message passing.

% \begin{figure}[t]
%     \centering
%     \includegraphics[width=0.8\linewidth]{figs/Message_Passing.pdf}
%     \vspace{-1mm}
%     \caption{Bidirectional (a) simultaneous and  (b) asynchronous message passing schemes. Simultaneous message passing updates all node messages in one go, while asynchronous message passing takes more than a single step to update all node messages. b(i) Shows the forward pass process for asynchronous message passing in the forward pass node 3 would be updated simultaneously by both predecessors, but we show it in parts for readability. b(ii) Shows the backward pass process for asynchronous message passing. The direction of the graph is changed and nodes 1 \& 2 are updated asynchronously. $\mathcal{A}$ represents the aggregate function that aggregates the messages of neighboring nodes to the current node.}
%     \vspace{-2mm}
%     \label{fig:message_passing}
% \end{figure}

\subsubsection{Simultaneous}
The more common approach in graph neural networks, specifically Graph Convolution Networks (GCN), is simultaneous message passing. In D-VAE, simultaneous message passing starts by taking the initial type of each node and sharing it with all its immediate neighbors (downstream neighbors if uni-directional). These messages are then combined to create a new localized representation of each node based on its type and its neighbors. The localization extends to additional neighbors by iteratively repeating this process.

\begin{figure*}[t]
    \centering
    \includegraphics[width=.9\linewidth]{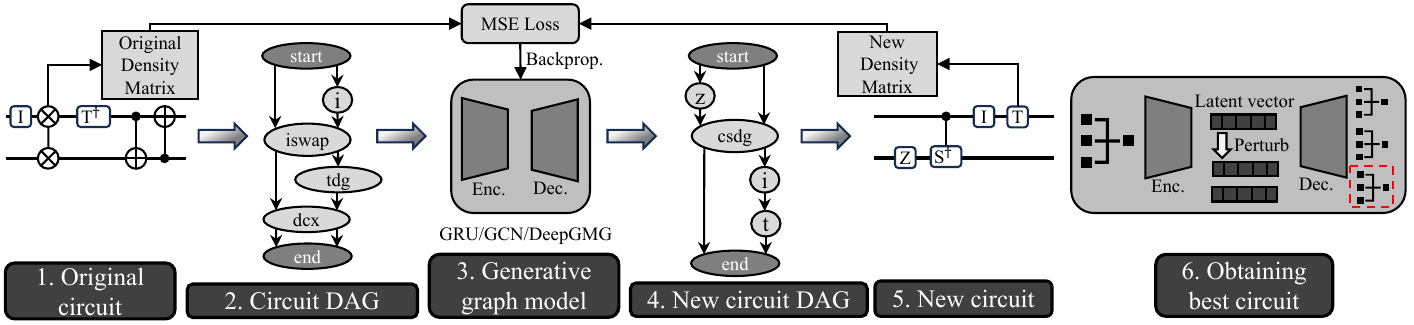}
    % \vspace{-6mm}
    \caption{Overview of AltGraph process for generating alternate representation of quantum circuits.}
    \label{fig:altgraph}
    \vspace{-2mm}
\end{figure*}

\subsubsection{D-VAE Asynchronous}
D-VAE proposes an ``asynchronous" message-passing scheme. This name may seem misleading since D-VAE does require some synchronization. Analysis of the first few gates of Fig. \ref{fig:sample_circ_graph}b is done for clarity.
At the top is the initial or the ``start" node. Once the representation of this node is created the children begin to process, however not all three children can process immediately. D-VAE requires processing all predecessor nodes before updating a node's state. So, only the ``csdg" and the ``y" gate may asynchronously process. The ``csx" will not be active until the completion of the ``cy" node following ``csdg". While the synchronization is time-consuming, it allows each node a ``complete" understanding of all predecessor nodes.

% Figure ~\ref{fig:sim_message}, contains an example of a simple network with a single iteration of the node state sharing process. 

% \begin{figure}[t]
% \begin{adjustbox}{width=\linewidth}
%   \begin{minipage}{\linewidth}
%      \centering
%      \includegraphics[width=.5\textwidth]{figs/sim_message.pdf}     
%   \end{minipage}\hfill
%   \end{adjustbox}
%   \caption{Sample simultaneous message passing.} \label{fig:sim_message}
% % \end{center}
% \end{figure}

\subsection{Generative Graph Models}
Generative graph models create new equivalent models based on input graphs and various constraints.
%; the architecture is malleable to generate quantum circuits. Generating equivalent quantum circuits requires converting a quantum circuit to a DAG and sending the DAG to the graph model. The graph model generates an equivalent DAG and translates it back to a quantum circuit. %There are several ways to ensure the equivalence of quantum circuits. In our architecture equivalence is evaluated using MSELoss, comparing the reconstructed circuit's density matrix to the original circuit's density matrix. 
D-VAE encodes a DAG to the latent space, then the encoding is used to obtain reconstructed DAGs. For detailed analysis, different variations of the D-VAE generative graph model \cite{zhang2019d} are used. Namely, D-VAE GRU, D-VAE GCN, and D-VAE DeepGMG \cite{li2018learning}. All three versions use topological ordering to create their DAGs, differing mainly in the encoding method.

\subsubsection{D-VAE GRU}
The standard D-VAE approach uses the Gated Recurrent Unit (GRU), which encodes the current node's state using the node's predecessors' states and the current node type. D-VAE uses asynchronous message passing \cite{li2015gated} and topological sorting to ensure all information is available to the node before encoding. D-VAE defaults to bi-directional encoding, where upon completion of forward encoding D-VAE reverses the graph and the node states are updated by repeating the same process.

\subsubsection{D-VAE GCN}
The second D-VAE approach uses GCN. D-VAE allows the designer to decide how many rounds of simultaneous message passing \cite{kipf2016semi} they desire to create the encodings of the nodes. To initialize the encoding D-VAE uses the neighbor (or predecessors if uni-directional) node types to initialize each node's state. In additional rounds of encoding neighbor (or predecessor) node states are used to update the state. Standard GCN models are primarily for encoding, so D-VAE applies the GRU method for GCN decoding.

\subsubsection{D-VAE DeepGMG}
The third D-VAE approach uses DeepGMG \cite{li2018learning}. The original DeepGMG work focuses on undirected graph generation. It also does not use topological ordering and uses a random order instead. To make a more valid comparison, D-VAE updates the design of DeepGMG to use topological ordering, allowing for multiple iterations of message passing and the processing of directed graphs. DeepGMG relies on a similar simultaneous message-passing scheme as GCN with an important update. DeepGMG, instead of only relying on surrounding node information, also accounts for an edge feature vector to better understand node connections.

% , GraphRNN \cite{you2018graphrnn},

\section{AltGraph: Proposed Approach} \label{altgraph}
%D-VAE is a powerful tool for processing DAGs and even shows it's ability to generate valid neural network architectures. 
\subsection{Approach}
To leverage the generative abilities of D-VAE for the quantum domain, we propose AltGraph, a robust approach to generate valid equivalent quantum circuits. In quantum graph representation, nodes represent quantum gates in a system, and edges represent the connection between the gates. We discuss the changes required in original D-VAE to realize AltGraph in the following paragraphs.

The overall process of generating quantum circuits using AltGraph is shown in Fig. \ref{fig:altgraph}. First, the original quantum circuits (step \raisebox{0.5pt}{ \textcircled{\raisebox{-0.9pt}{1}}}) are converted into DAGs (step \raisebox{0.5pt}{ \textcircled{\raisebox{-0.9pt}{2}}}). These DAGs are sent to the generative graph model (step \raisebox{0.5pt}{ \textcircled{\raisebox{-0.9pt}{3}}}) for reconstructing new DAGs (step \raisebox{0.5pt}{ \textcircled{\raisebox{-0.9pt}{4}}}), from which the new quantum circuit is reconstructed (step \raisebox{0.5pt}{ \textcircled{\raisebox{-0.9pt}{5}}}). For training, we use MSE Loss between the density matrices of original and new quantum circuits to ensure similar functionality. Finally, to obtain the best quantum circuit, we perturb the latent representation of the circuit with some noise (step \raisebox{0.5pt}{ \textcircled{\raisebox{-0.9pt}{6}}}) and obtain multiple candidate circuits. From these, one can select the circuit with matching density metrix and least depth and gate count.

\subsection{Loss}
D-VAE used teacher forcing to get faster convergence of models. For loss evaluation, D-VAE used encoder KLD and the model's ability to predict the correct node. D-VAE also evaluated the loss based on the model's ability to predict edges using the ``true" node. While these features are viable when dealing with graphs and ensuring similarly designed graphs, the goal is to generate equivalent quantum circuits. Structurally similar graphs are not a guarantee of equivalent quantum circuits. It's necessary to compare density matrices to evaluate quantum circuit equivalence. The loss process is updated to generate the nodes and edges rather than forcing edge predictions based on the ``true" node. This forces complete graph reconstruction for each input, allowing density matrix comparison. The density matrices are taken from the target hardware post-transpilation using Qiskit's default optimization level 1. This ensures the circuits are compared based on their real values rather than the ideal state.
AltGraph's loss calculation uses the KLD of the encoder, the mean squared error (MSE) of the original circuit's density matrix, and the reconstructed circuit's density matrix.

\subsection{Graph Generation}
D-VAE requires that each graph has a single input node and a single output node. In cases with multiple input/output nodes, D-VAE requires a virtual node to represent the final input/output. To match D-VAE requirements, input qubit wires map to the start node, and all qubit measurements map to the end node. An example mapping is in Fig. ~\ref{fig:sample_circ_graph}.

Quantum circuits have numerous requirements to ensure validity. Ideally, a neural network could learn the basic requirements of quantum circuit structures. However, the loss calculations use the difference between the density matrices of the original and reconstructed circuit, so even the initial circuits need to be valid constructed designs.

\subsubsection{Number of Qubits}
The first requirement for density matrix comparison is that both density matrices are the same size. Density matrices are size $2^{N} * 2^{N}$ where $N$ is the number of qubits. To ensure reconstructed circuit size equivalence, the reconstructed DAG start node out-degree must match the out-degree of the original DAG's start node. The reconstructed DAG end node must have the same in-degree as the original DAG's end node. 

\subsubsection{Gates \& Gate Connections}
D-VAE uses a topological ordering for both encoding and reconstruction. During reconstruction, it allows for the selection of edges between the current node and any previous node. D-VAE also allows the model to learn the appropriate amount of in and out edges. However, quantum gates have a specified in and out edge limitation. To ensure the use of all out-edges, the model maintains a queue of available nodes for the current node to connect. The node is removed from the queue when the out-edges are complete. When the model predicts the ``end" node it connects all remaining nodes in the queue. The model predicts the most likely connection from the available list of previous nodes for the in edge. If it is a multiple-wire gate, the prediction process repeats until all in-degrees are filled. 

\subsubsection{Circuit Types} \label{circ_types}
To reduce the search space complexity of quantum circuits training/testing uses graphs made of non-parametric circuits that consist of one and two-qubit gates. Specifically, the single qubit gates $x$, $y$, $z$, $h$, $s$, $t$, $id$, $sxdg$, $sdg$, $sx$, $tdg$ and $cx$, $cy$, $cz$, $swap$, $dcx$, $iswap$, $csdg$, $ecr$, $ch$, $cs$, $csx$ two-qubit gates are used.

\section{Results} \label{results}

% \subsection{Setup}
% While generating equivalent quantum circuits, the maximum limit of gate predictions is set to the original circuit gate count. While this may result in higher loss values for some circuits than is achievable, it punishes the model for creating larger circuits. The additional punishment that occurs also addresses the goal of creating more hardware-compliant circuits. To evaluate AltGraph's performance, the qubit count sweeps between 2, 4, and 6 qubits, and the gate count to 16, 24, and 32 gates per qubit count. A random circuit generator produces 300 random graphs for each quantum circuit size using the gates given in Section \ref{circ_types}. The training uses a 90-10 training/testing split, resulting in 270 training sample circuits and 30 testing circuits. The same 300 samples are utilized for every model, eliminating possible discrepancies that might result from varying available data samples. Per each circuit size per model, the model trains three separate times. All values reported are the mean value of the results.

\subsection{Setup and Evaluation Framework}
In generating equivalent quantum circuits, the maximum gate predictions are capped at the original circuit gate count. This can yield higher loss values but penalizes models for larger circuits, aligning with the aim of more hardware-friendly circuits. AltGraph's performance is assessed with qubit counts of 2, 4, and 6 and gate counts of 16, 24, and 32 per qubit. A total of 300 random circuits for each size are produced using gates from Section \ref{circ_types}. The data follows a 90-10 training/testing split. The same 300 samples are used for all models, ensuring consistency. Each model is trained thrice per circuit size, and reported values represent the average outcomes.

% \begin{figure*}[t]
% \begin{adjustbox}{width=\linewidth}
%   \begin{minipage}{\linewidth}
%      \centering
%      \includegraphics[width=\linewidth]{figs/qc_loss.pdf}     
%   \end{minipage}\hfill
%   \end{adjustbox}
%   \caption{AltGraph averaged loss per quantum circuit size.} \label{fig:loss}
% % \end{center}
% \end{figure*}

\begin{table}[t]
\vspace{-4mm}
\caption{Average gate reduction (\%) and loss per model per gate count.}
\vspace{-2mm}
\label{tab:loss}
\begin{adjustbox}{width=\linewidth}
\begin{tabular}{@{}c|cccccc@{}}
\toprule
 & \multicolumn{2}{c}{2 Qubit} & \multicolumn{2}{c}{4 Qubit} & \multicolumn{2}{c}{6 Qubit} \\ \midrule
\textbf{Model Type \#Gate} & \textbf{Loss} & \textbf{Reduction} & \textbf{Loss} & \textbf{Reduction} & \textbf{Loss} & \textbf{Reduction} \\
GRU 16 & 0.09 & 19.59\% & 0.01 & 24.63\% & 0.01 & 30.18\% \\
GRU 24 & 0.09 & 33.94\% & 0.01 & 43.86\% & 0.01 & 43.96\% \\
GRU 32 & 0.10 & 36.27\% & 0.01 & 51.37\% & 0.01 & 54.16\% \\
GCN 16 & 0.22 & 19.83\% & 0.12 & 26.17\% & 0.11 & 32.29\% \\
GCN 24 & 0.31 & 33.02\% & 0.22 & 41.09\% & 0.19 & 41.99\% \\
GCN 32 & 0.54 & 35.34\% & 0.44 & 51.74\% & 0.43 & 53.1\% \\
DeepGMG 16 & 0.14 & 19.57\% & 0.03 & 23.67\% & 0.06 & 30.34\% \\
DeepGMG 24 & 0.18 & 31.99\% & 0.10 & 31.71\% & 0.08 & 42.6\% \\
DeepGMG 32 & 0.28 & 40.77\% & 0.11 & 40.44\% & 0.12 & 51.50\% \\ \bottomrule
\end{tabular}
\end{adjustbox}
\vspace{-6mm}
\end{table}

\subsection{Results and Analysis}
Table \ref{tab:loss} presents the mean training loss and gate reduction percentages for three models: AltGraph GRU, AltGraph GCN, and AltGraph DeepGMG across various circuit sizes. These percentage reductions are derived from comparing the number of gates between the transpiled original and the transpiled reconstructed circuits. % calculated using Equation ~\ref{eq:reduction}.

%\begin{equation} \label{eq:reduction}
%\%Reduced = \frac{\#gates_{O} - \#gates_{R}}{\#gates_{O}}
%\end{equation}

The GRU model's loss tends towards zero faster than both the GCN and DeepGMG models, irrespective of the qubit and gate counts. Moreover, DeepGMG's loss approaches zero faster than GCN. Interestingly, despite GCN's generally higher loss, it surpasses both GRU and DeepGMG when circuits have 16 gates, irrespective of the qubit count. Conversely, for circuits with 24 gates, GRU is superior, irrespective of qubit count. For 32-gate circuits, performance varies by qubit count: DeepGMG excels for 2-qubit circuits, GCN for 4-qubit circuits, and GRU for 6-qubit circuits.

\begin{figure*}[t]
\begin{adjustbox}{width=\linewidth}
  \begin{minipage}{\linewidth}
     \centering
     \includegraphics[width=0.9\linewidth]{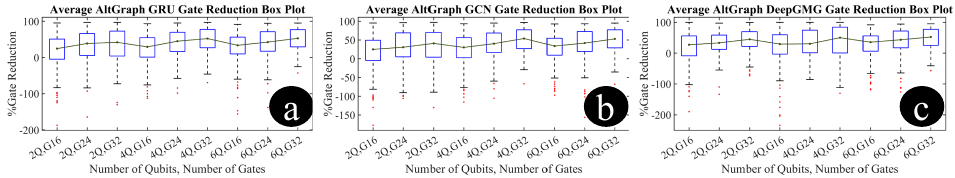}     
  \end{minipage}\hfill
  \end{adjustbox}
  \caption{Averaged gate reduction per circuit size for (a) AltGraph GRU, (b) AltGraph GCN,(c) AltGraph DeepGMG.} \label{fig:box-gate}
  \vspace{-2mm}
% \end{center}
\end{figure*}

Delving deeper into each model's capability to yield a circuit with fewer gates post-transpilation provides further insights into their efficiencies. This analysis involves three different encodings for each testing gate and three decodings for each encoding sample, resulting in nine reconstructions per circuit. Upon generating these samples, both the reconstructed and original circuits are transpiled using Qiskit's \textit{FakeWashingtonV2} to emulate performance on cutting-edge hardware. Fig. \ref{fig:box-gate}a showcases the mean gate reduction for the AltGraph GRU model per transpiled circuit size.

\begin{figure*}[t]
\begin{adjustbox}{width=\linewidth}
  \begin{minipage}{\linewidth}
     \centering
     \includegraphics[width=0.9\linewidth]{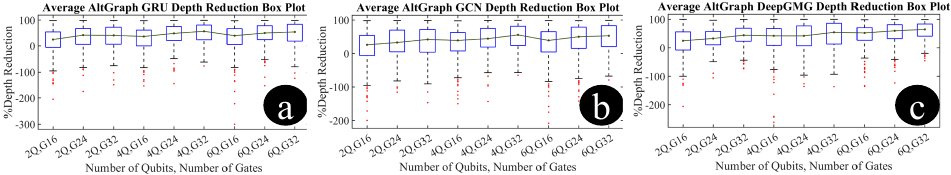}     
  \end{minipage}\hfill
  \end{adjustbox}
  \caption{Averaged depth reduction per circuit size for (a) AltGraph GRU, (b) AltGraph GCN, (c) AltGraph DeepGMG.} \label{fig:box-depth}
  \vspace{-4mm}
% \end{center}
\end{figure*}

% \color{blue}From Fig. \ref{fig:box-gate}a, we note a net positive percentage gate reduction for all varying qubits and gates. We observe a median gate count reduction ranging from 24\% (2 qubits, 16 gates) to nearly 53\% (6 qubits, 32 gates). For each qubit count, as the number of gates increases, the subsequent percentage of gate reduction also increases. For example, the median gate reduction for 6 qubit circuits with 16, 24, and 32 gates are 33.6\%, 42.4\%, and 52.9\% respectively. A similar trend is observed for 2 and 4-qubit circuits as well. There are also a few instances of generated quantum circuits with increased gate counts. For example, there is an instance of a 4 qubit, 16 gate quantum circuit which when reconstructed has a nearly 73\% increase in gate count. However, such instances are rare and are mostly outliers. As can be seen from Fig \ref{fig:box-gate}a, most of the alternate representations generated lead to an overall decrease in the number of gates. \color{black}
\textbf{Gate Reduction Trends Across Models:} From Fig. \ref{fig:box-gate}a, a consistent positive gate reduction is observed across different qubits and gate configurations. Median gate reductions range from 24\% (2 qubits, 16 gates) to 53\% (6 qubits, 32 gates). As the gate count increases within each qubit category, there's a proportional increase in gate reduction percentages. Specifically, for 6 qubit circuits with 16, 24, and 32 gates, the median reductions stand at 33.6\%, 42.4\%, and 52.9\%, respectively. Comparable patterns are seen for 2 and 4-qubit circuits. There are few reconstructions that result in an increased gate count, like a 4 qubit, 16 gate circuit showing a surge of 73\%. However, such cases are exceptions, as Fig. \ref{fig:box-gate}a indicates that most alternative configurations yield a reduction in gate count.

Trends seen in GCN and DeepGMG match those of GRU, as demonstrated in Figures \ref{fig:box-gate}b and \ref{fig:box-gate}c. GCN's median gate reduction spans from 25\% (2 qubits, 16 gates) to 54\% (4 qubits, 32 gates), and DeepGMG's range is from 27\% (2 qubits, 16 gates) to 52.4\% (6 qubits, 32 gates). For both models, as the number of gates grows, gate reduction percentages correspondingly increase. Comparatively:
\raisebox{0.5pt}{ \textcircled{\raisebox{-0.9pt}{1}}} For 2-qubit circuits, DeepGMG achieves the peak in median gate reduction, followed by GRU and then GCN, 
\raisebox{0.5pt}{ \textcircled{\raisebox{-0.9pt}{2}}} in 4-qubit circuits, GCN leads in median gate reduction, with GRU second and DeepGMG last, and 
\raisebox{0.5pt}{ \textcircled{\raisebox{-0.9pt}{3}}} for 6-qubit circuits, all models display similar median reductions, though DeepGMG has an advantage.
\textbf{Depth Reduction Trends Across Models:} From Fig. \ref{fig:box-depth}a, there's a clear and consistent depth reduction across different configurations of qubits and gates. The median depth reduction spans from 24\% for circuits with 2 qubits and 16 gates to 56\% for those with 4 qubits and 32 gates. Intriguingly, circuits with an identical qubit count display an enhanced depth percentage reduction with the addition of more gates. For 6-qubit circuits, the median depth reductions are 39.6\% for 16 gates, 49.9\% for 24 gates, and 53.9\% for 32 gates. This pattern is consistent for 2 and 4-qubit circuits as well.
While there are rare quantum circuits that show an increased depth post-reconstruction, such as a 4-qubit, 16-gate circuit with a 100\% depth increase, these cases are rare, predominantly outliers. As illustrated in Fig. \ref{fig:box-depth}a, the dominant trend is a depth reduction in the alternative representations.

GCN and DeepGMG trends reflect those of GRU, as observed in Figures \ref{fig:box-depth}b and \ref{fig:box-depth}c. GCN's median depth reduction ranges from 26\% (2 qubits, 16 gates) to 55\% (4 qubits, 32 gates), while for DeepGMG, it's between 24\% (2 qubits, 16 gates) and 58\% (6 qubits, 32 gates). As gate counts rise for every qubit configuration, depth reduction percentages also climb for both GCN and DeepGMG. Comparative analysis of the models reveals:
\raisebox{0.5pt}{ \textcircled{\raisebox{-0.9pt}{1}}} For 2-qubit circuits, GRU tops in median depth reduction, succeeded by DeepGMG and then GCN.
\raisebox{0.5pt}{ \textcircled{\raisebox{-0.9pt}{2}}} With 4-qubit circuits, GRU retains the lead in median depth reduction, followed by GCN and subsequently DeepGMG.
\raisebox{0.5pt}{ \textcircled{\raisebox{-0.9pt}{3}}} In 6-qubit circuits, DeepGMG stands out with the highest median depth reduction, with GCN in the second spot and GRU trailing.

\begin{table}[t]
  \caption{Average gate reduction (\%) per model per gate count}
  \vspace{-4mm}
  \label{tab:average-pgc-reduction}
  \begin{center}
  % \begin{adjustbox}{width=\linewidth}
        \begin{tabular}{c|ccc}
            \toprule
            \textbf{Model} & \textbf{16 Gate} & \textbf{24 Gate} & \textbf{32 Gate} \\
            \hline
            AltGraph GRU & 24.80\% & 40.59\% & 47.27\%  \\
            % \hline
            AltGraph GCN & 26.10\% & 38.70\% & 46.73\% \\
            % \hline
            AltGraph DeepGMG & 24.53\% & 35.44\% & 44.24\% \\
            \bottomrule
        \end{tabular}
    \end{center}
    \vspace{-2mm}
  % \end{adjustbox}
\end{table}

% Examining Table ~\ref{tab:average-pgc-reduction}, AltGraph GRU effectively reduces gate count compared to both the GCN and DeepGMG models for the 24 and 32 gate circuits. In the case of 16 gates the GCN model performs better with a 5.11\% percent difference between the GCN and the GRU model. Table ~\ref{tab:average-reduction} tabulates the overall gate reduction performance per model.

\begin{table}[t]

  \caption{Average gate reduction per model}
  
  \vspace{-4mm}
  \label{tab:average-reduction}
  \begin{adjustbox}{width=.8\linewidth}
  % \begin{center}
        \begin{tabular}{c|cc}
            \toprule
            \textbf{Model} & \textbf{Average Gates Reduced} & \textbf{Average Reduction\%} \\
            \hline
            AltGraph GRU & 35.67 & 37.55 \\
            % \hline
            AltGraph GCN & 33.78 & 37.17 \\
            % \hline
            AltGraph DeepGMG & 32.87 & 34.73 \\
            Qiskit & 26.15 & 18.65 \\
            t$\ket{ket}$ & 73.68 & 33.5 \\
            Quartz \cite{xu2022quartz} & -- & 30.1 \\
            Quarl \cite{li2023quarl} & -- & 36.6 \\
            Monte Carlo Tree \cite{zhou2020monte} & -- & 30.0 \\
            \bottomrule
        \end{tabular}
  % \end{center}
  \end{adjustbox}
  \vspace{-2mm}
\end{table}

% Examining Table ~\ref{tab:average-reduction}, AltGraph GRU averages a higher reduction of gates than the GCN or DeepGMG model, but GCN remains competitive with only a 1.02\% percent difference in performance to the GRU model. Table ~\ref{tab:average-depth} tabulates the overall depth reduction performance per model.
\textbf{Comparative Analysis of Model Performances:} 
\raisebox{0.5pt}{ \textcircled{\raisebox{-0.9pt}{1}}}
From Table \ref{tab:average-pgc-reduction}, AltGraph GRU stands out in its gate count reduction capabilities when compared to both GCN and DeepGMG for circuits with 24 and 32 gates. However, for 16-gate circuits, the GCN model surpasses, showing a 5.11\% performance difference over the GRU model. The cumulative gate reduction performance of each model is detailed in Table \ref{tab:average-reduction}. \raisebox{0.5pt}{ \textcircled{\raisebox{-0.9pt}{2}}}
Upon inspecting Table \ref{tab:average-reduction}, it's evident that the AltGraph GRU, on average, trims more gates than either the GCN or DeepGMG models. Yet, GCN remains a strong contender, lagging behind GRU by a mere 1.02\%. To understand how AltGraph performs compared to state-of-the-art systems we include the results from optimizing the same circuits using two rules based methods, t$\ket{ket}$ and Qiskit. We also include the reported average improvements of 3 state-of-the-art systems ran on IBM Qiskit non-parametric circuits. \cite{xu2022quartz, schulman2017proximal, zhou2020monte}.

The collective depth reduction performance for each model is compiled in Table \ref{tab:average-depth}. \raisebox{0.5pt}{ \textcircled{\raisebox{-0.9pt}{3}}}
A closer look at Table \ref{tab:average-depth} reveals that the AltGraph DeepGMG consistently outperforms in depth reduction when set against the GCN and GRU models. Nevertheless, the GRU model isn't far behind, showing only a slight 5.98\% performance gap to the DeepGMG. For a comprehensive view of the average MSE for test circuits, refer to the last column of Table \ref{tab:average-depth}. \raisebox{0.5pt}{ \textcircled{\raisebox{-0.9pt}{4}}}
Finally, analyzing Table \ref{tab:average-depth}, it becomes clear that AltGraph GRU excels in achieving a smaller density matrix difference than both the GCN and DeepGMG models.

% \begin{table}[t]
%   \caption{Average depth reduction per model.}
%   \vspace{-2mm}
%   \label{tab:average-depth}
%   \begin{center}
%         \begin{tabular}{|c|c|}
%             \hline
%             \textbf{Model} & \textbf{Average Depth Reduction\%} \\
%             \hline
%             AltGraph GRU & 37.75 \\
%             \hline
%             AltGraph GCN & 35.54 \\
%             \hline
%             AltGraph DeepGMG & 40.08 \\
%             \hline
%         \end{tabular}
%   \end{center}
%   \vspace{-4mm}
% \end{table}
\begin{table}
  \caption{Average depth reduction and test circuit MSE per model}
  \vspace{-4mm}
  \label{tab:average-depth}
  \begin{adjustbox}{width=.8\linewidth}
  % \begin{center}
        \begin{tabular}{c|cc}
            \toprule
            \textbf{Model} & \textbf{Average Depth Reduction\%} & \textbf{Average MSE} \\
            \hline
            AltGraph GRU & 37.75 & .0074 \\
            % \hline
            AltGraph GCN & 35.54 & .27 \\
            % \hline
            AltGraph DeepGMG & 40.08 & .13 \\
            \bottomrule
        \end{tabular}
  % \end{center}
  \end{adjustbox}
  \vspace{-4mm}
\end{table}

\section{Discussion} \label{discussion}

%This work is intended to be a proof-of-concept work that does not rely on quantum hardware compliance, such as coupling map and native gate set, and focuses on non-parametric quantum circuits.
\subsection{Mixed Optimization Techniques}
Drawing from Table \ref{tab:average-pgc-reduction}, it's evident that graph generative models present a promising avenue for gate reductions across different circuit dimensions. However, there's scope for refining the models further. The circuits used in this study were randomly generated and didn't undergo optimization using conventional rule-based methodologies. If models were trained on optimized circuits, it's plausible that there would be a more significant overall reduction between original and reconstructed circuits, as the model would target only substitutions not already addressed. Implementing rule-based optimization post circuit reconstruction could yield added reductions.

\subsection{MSE Constraints}
Despite the models exhibiting robust learning capabilities, using MSELoss as an evaluative metric has its drawbacks. While MSELoss serves well in gauging matrix similarity, there are pitfalls in its application. Being a regressive optimizer, MSELoss doesn't always ensure the discovery of the best solution. In the quantum realm, this can translate to functional discrepancies. For instance, if a model, during training, predicts a Pauli-X gate  
($\begin{bsmallmatrix}
0 & 1\\
1 & 0
\end{bsmallmatrix}$) 
for an original Pauli-Z gate
($\begin{bsmallmatrix}
1 & 0\\
0 & -1
\end{bsmallmatrix}$) 
, a high loss results. But if the model later predicts an S gate ($\begin{bsmallmatrix}
  1 & 0\\
  0 & i
\end{bsmallmatrix}$) 
, the resulting lower loss might misleadingly suggest an ``optimized'' circuit, which isn't functionally accurate. A deeper probe into the relationship between a low MSELoss and a model's functional accuracy is warranted.

\subsection{Density Matrix Constraints}
Another challenge with MSELoss arises with sparse density matrices. In cases with notably high qubit counts and minimal gate counts, MSELoss might fall short. For illustration, consider a 16 qubit circuit with ($2^{16} * 2^{16} = 4,294,967,296$) values that possesses only one single qubit gate. Here, the MSELoss for a circuit with any such gate would be nearly zero. Yet, this limitation can be overlooked as the primary objective is to optimize large, intricate circuits, rendering sparse configurations, like the one mentioned, irrelevant to the study's context.

Scalability to larger qubit counts initially seems problematic, as density matrices suffer from Nth power scaling. However, scaling can be addressed by performing block replacements of sub-circuits. For example, if there is a 100 qubit quantum circuit, the density matrix will have $2^{100}$ elements, which will be difficult to process. However, one can break it down into blocks of five qubit sub-circuits and generate alternate representations for each sub-circuit one by one. 

\section{Conclusions} \label{conclusion}
In our study, generative graph models have proven to be a promising method for gate reduction in quantum circuits. The AltGraph GRU model showcased an average gate count reduction of 37.55\%, coupled with a depth reduction of 37.75\%, and achieved an impressively low average MSE of 0.0074. The AltGraph GCN model delivered a comparable gate count reduction of 37.17\%, a depth reduction of 35.54\%, but with a slightly higher average MSE of 0.27. Finally, the AltGraph DeepGMG model, while averaging a gate reduction of 34.73\% and achieving the highest depth reduction of 40.08\%, reported an average MSE of 0.13. These findings underscore the potential and versatility of generative graph models in the realm of quantum circuit optimization.

\bibliographystyle{ACM-Reference-Format}
\bibliography{sample-base}

% \appendix

\end{document}